\documentclass[12pt]{iopart}
\usepackage{iopams}
\usepackage{graphicx}
\newcommand{\be}{\begin{equation}}
\newcommand{\ee}{\end{equation}}
\newcommand{\bef}{\begin{figure}}
\newcommand{\eef}{\end{figure}}
\newcommand{\bea}{\begin{eqnarray}}
\newcommand{\eea}{\end{eqnarray}}
\newcommand{\bx}{{\bf x}}
\newcommand{\bk}{{\bf k}}
\newcommand{\bq}{{\bf q}}

\newcommand{\by}{{\bf y}}

\newcommand{\bu}{{\bf u}}
\newcommand{\bv}{{\bf v}}
\newcommand{\bw}{{\bf w}}
\newcommand{\la}{\left<}
\newcommand{\ra}{\right>}

\begin{document}
\title{Diffusion, super-diffusion and coalescence from single step} 

\author{Andrea Gabrielli, Fabio Cecconi} 

\address{
SMC, INFM-CNR, Department of Physics,\\
University ``La Sapienza'' of Rome, P.le Aldo Moro 2, 00185-Rome, Italy;\\
ISC-CNR, via dei Taurini 19, 00185-Rome (Italy)}

\hyphenation{Lan-ge-vin}


\begin{abstract}
From the exact single step evolution equation of the two-point 
correlation function of a particle distribution subjected to a 
stochastic displacement field $\bu(\bx)$, we derive different dynamical regimes
when $\bu(\bx)$ is iterated to build a velocity field.
First we show that spatially uncorrelated fields $\bu(\bx)$ lead to
both standard and anomalous diffusion equation.  When the field
$\bu(\bx)$ is spatially correlated each particle performs a simple
free Brownian motion, but the trajectories of different particles 
result to be mutually correlated. 
The two-point statistical properties of the field $\bu(\bx)$
induce two-point spatial correlations in the particle distribution
satisfying a simple but non-trivial diffusion-like equation.  These
displacement-displacement correlations lead the system to three
possible regimes: coalescence, simple clustering and a combination of
the two. The existence of these different 
regimes, in the one-dimensional system, is shown through computer 
simulations and a simple theoretical argument. 
\end{abstract}

\submitto{Journal of Statistical Mechanics: Theory and Experiments}

\maketitle

\section{Introduction} \label{intro}
The study of diffusion phenomena has a long history dating back to
decades since the origin of statistical mechanics. It includes many
different subjects and applications ranging from irregular motion of
particles in homogeneous or disordered media and osmosis
\cite{Disorder,Osmosis}, to standard and anomalous transport
\cite{Transport,George} of heat, mass and charges in materials, from
coalescence of passive scalars \cite{Scalars} in highly turbulent
fluxes to biological and ecological investigations on animal dispersal
\cite{Murray,Levin}.

The basis of the theory of diffusion is the random walk (RW) or random
flight where a particle undergoes successive random displacements from
its initial positions performing an irregular motion. The resulting
character of the particle trajectory emerges from the statistical
properties of these displacements. The wide applicability of the RW to
natural phenomena relies just on the possibility to introduce
appropriate generalizations on probabilistic nature of the
displacements. One of the straightforward generalization is realized
by introducing correlations in the displacements so to obtain the so
called correlated random walks (CRW) \cite{CRW}.  This possibility
extends also to a set of particles distributed in space leading to the
definition of spatially correlated random walks.  In this case one can
wonder what kind of particle distribution emerges from reiterated
displacements of the particles and how its properties can be directly
inferred from the knowledge of the statistical correlation of the
displacements \cite{Melig,coha}.  This approach is usually
adopted to model the motion of a system of particles in a disordered
environment where the randomness change in time. The study of this
kind of models is strictly related to the mathematical theory of the
so-called point-processes \cite{PointP}, i.e., stochastic point-wise
particle distributions. The theory which investigates the geometrical 
and statistical properties of the stochastic processes generating particle
distributions constitutes a very important
field of research with important physical and cross-disciplinary
applications \cite{Astrophys}.  

In this paper we derive different results on the physics of
diffusion  
from the exact transformation \cite{displa} of the
two-point correlations of a particle distribution subjected to a
single step stochastic deformation. More precisely,
we deal with the transformation that a stochastic displacement field
induces on given particle distribution once it is applied to the 
particles. Statistically
independence between the particle distribution and the displacement
field is assumed, but arbitrary displacement-displacement correlations
and initial density-density correlations can be present.  The main
purpose of this study was originally to understand, through the exact
formula presented for one a two-point particle correlations in
\cite{displa}, the discretization and finite size effects in the
preparation of the initial conditions in cosmological $n-$body 
\cite{frenk,bruno}
simulations which are usually built by applying a suitable stochastic
deformation field to a regular lattice or other very uniform
configurations of particles with equal mass.  In the present context,
instead, we use the same approach to derive by reiteration
all the famous equations of ordinary diffusion, super and
sub-diffusion. We finally consider the most complex case of clustering and
coalescence of particles observed in fully developed turbulence due
to spatially correlated diffusion induced by turbulent flows on 
passive pollutants.

\section{From one step displacement to spatially correlated random walks}
\label{sec1}

In this section we recall first some notations for the one and
two point correlation function of a spatially homogeneous point
process, then we briefly summarize the main results of Ref.~\cite{displa}.

Let us consider a $d-$dimensional spatial distribution of $N$ particles 
with equal unitary mass (i.e. a so called ``point process'') in a volume
$V$ which we assume to coincide asymptotically with $I\!\!R^d$. 
The microscopic density is by definition
$$
n(\bx)=\sum_{i=1}^{N}\delta(\bx-\bx_i)\,,
$$ 
where the limit is taken keeping fixed $N/V=n_0$, and $\bx_i$
indicates the position of the $i^{th}$ particle.  The system is
supposed to be statistically homogeneous and uniform on large scales,
therefore the average density is well defined and positive: $ \la
n(\bx)\ra = n_0>0$, where the average $\la ..\ra$ runs over the
ensemble of realizations of the point process.  Moreover, we define
the normalized connected two-point correlation function, as \be
\xi(\bx)=\frac{\la n(\bx_0) n(\bx_0+\bx)\ra }{ n_0^2}-1=
\frac{\delta(\bx)}{ n_0}+h(\bx)\,.
\label{eq1}
\ee
The {\em covariance} function $h(\bx)$ is the off-diagonal two-point correlation function
and gives the true correlation between different particles.
The power spectrum of the particle distribution is defined 
as the limit
$$
S(\bk)=\lim_{V\rightarrow\infty}\frac{\left<|\hat n(\bk;V)|^2\right>}{N}
-2\pi n_0\delta(\bk)\;,
$$
where 
$
\hat n(\bk;V)=\int_Vd^3x\,\exp(-i\bk\cdot\bx) n(\bx)
$ 
is the  Fourier transform (FT) of the density on the finite volume $V$. 
Because of the hypothesis on statistical homogeneity, $S(\bk)$ is simply 
given by the Fourier transform (FT) of $\xi(\bx)$ multiplied by $ n_0$:
\be
S(\bk)= n_0{\cal F}[\xi(\bx)]=1+ n_0 \hat h(\bk)\,,
\label{eq2}
\ee 
with $\hat h(\bk)={\cal F}[h(\bx)]$ and ${\cal F}[..]=\int d^dx\,
e^{-i\bk\cdot\bx}(..)$ 
being the usual $d-$dimensional infinite volume Fourier transform. 
It is clear that for a
statistically homogeneous particle distribution, $S(\bk)$ and $\xi(\bx)$ 
contains the same information. 

When a statistically homogeneous and arbitrarily correlated stochastic
displacement field $\bu(\bx)$ is applied to the particle distribution,
each particle moves from its old position $\bx_i$ to the new one
$\bx_i+\bu(\bx_i)$.  We assume that the displacement field and the
particle positions are statistically independent. We are interested in
the change of the two-point correlation properties of the particle
distribution under the effect of the displacements.  The complete
statistics of $\bu(\bx)$ is given by a
probability density functional ${\cal P}[\bu(\bx)]$ giving the
statistical weight of each realization of the stochastic
field. However, in our hypotheses, the behavior
of the two-point correlation function or the power spectrum under the system displacements \cite{displa}
is only determined through the knowledge of the probability density 
function (PDF) 
\be
\phi(\bw;\bx)=\int\!\int
d^du\,d^dv\,f(\bu,\bv;\bx)\delta(\bw-\bu+\bv)\,.
\label{phi}
\ee
that two particles, separated by the vector distance $\bx$, undergo a
relative displacement $\bw$. In Eq.~(\ref{phi}) we used the properties 
that $\phi(\bw;\bx)$ is, in turn, related to the joint PDF
$f(\bu,\bv;\bx)$ that two particles separated by the vector $\bx$
perform the displacements $\bu$ and $\bv$ respectively. 
If we denote by $S_{in}(\bk)$ and 
$\xi_{in}(\bx)= [\delta(\bx)/n_0+h_{in}(\bx)]$ respectively the 
power spectrum and the two-point correlation function 
of the particle distribution before the application of
the displacements and by 
$S_{f}(\bk)$ and 
$\xi_{f}(\bx)=[\delta(\bx)/ n_0+h_{f}(\bx)]$ the corresponding
quantities after the displacements,
we can write \cite{displa} the equation:
\be
S_f(\bk)=1-\int d^dq\,\tilde\phi(\bk,\bq)+
 n_0\int d^dx\, e^{-i\bk\cdot\bx}\hat\phi(\bk;\bx)[1+\xi_{in}(\bx)]
-(2\pi)^d n_0\delta(\bk)\,,
\label{eq3}
\ee
where 
\[\hat\phi(\bk;\bx)=\int d^dw\,e^{-i\bk\cdot\bw}\phi(\bw;\bx)\]
is the characteristic function\footnote{From Eq.~(\ref{phi}), it 
is immediate to verify also that
$\hat\phi(\bk;\bx)=\hat f(\bk,-\bk;\bx)$ and $\tilde\phi(\bk,\bq)=\tilde
f(\bk,-\bk;\bq)$ where respectively $\hat f(\bk,\bk';\bx)=
\int\!\int d^du\,d^dv\,f(\bu,\bv;\bx)e^{-i(\bk\cdot\bu+\bk'\cdot\bv)}$
and $\tilde f(\bk,-\bk;\bq)=\int d^dx\,\hat f(\bk,-\bk;\bx)e^{-i\bq\cdot\bx}$.}
of the random displacement $\bw$,
and $\tilde\phi(\bk,\bq)=\int d^dx\,e^{-i\bq\cdot\bx}\hat\phi(\bk;\bx)$.
The only hypotheses for the validity of Eq.~(\ref{eq3}) are:
(i) spatial homogeneity of both particle distribution and  
displacement field, (ii) statistical independence between the
particle positions and the displacement field.
Notice that no ``small displacements'' approximation or special properties 
of the displacement correlations are required.
Let us define 
\be 
G_{\mu\nu}(\bx)= \overline{u_{\mu}(\bx_0)u_{\nu}(\bx_0+\bx)}
\quad\quad\quad\quad \mu,\nu=1,...,d
\label{eq:G}
\ee 
the displacement-displacement correlation function which is
a symmetric tensor of rank $2$, whose FT,  is a non-negative 
definite symmetric tensor for all the $k$-vectors. 
With $\overline{(..)}$ indicating the average over ${\cal
P}[\bu(\bx)]$.  For symmetric distributions ${\cal
P}[\bu(\bx)]= {\cal P}[-\bu(\bx)]$, the positive value
$G_{\mu\mu}({\bf 0})$ represents the variance of the $\mu$-th
component of $\bu(\bx)$ at any point. 
Moreover in the case of a Gaussian field,  
the function $G_{\mu\nu}(\bx)$ determines completely  
the probabilistic properties of the field~\cite{Astrophys}.

The purpose of this paper is to deduce from Eq.~(\ref{eq3}) the
time-evolution of the two-point correlation function and 
power spectrum of a particle distribution in the
limit that each particle performs a Brownian trajectory, but the
motions of different particles can be arbitrarily spatially
correlated.  In other words, we consider an assigned
displacement field statistics, $f(\bu,\bv;\bx)$, at every time step
$\Delta t$, but no time correlation between
consecutive time-steps. Then we take $\Delta t\rightarrow 0$
in such a way to have a well defined diffusional limit.  Two cases
have to be basically distinguished:
\begin{enumerate}
\item 
The field $\bu(\bx)$ is a spatially
uncorrelated stochastic process at each time step, 
i.e., $\bu(\bx)$ and $\bu(\by)$ are completely independent 
if $\bx\ne \by$.  In this case,
$$
f(\bu,\bv;\bx)=
\left\{
\begin{array}{ll}
p(\bu)p(\bv)          & \mbox{if $\bx\neq 0$} \\
\delta(\bu-\bv)p(\bu) & \mbox{if $\bx=0$}
\end{array}\right.
$$
where 
$p(\bu)$ indicates the single displacement PDF \cite{displa}.  
Basically, the displacement field is simply a white noise both in
space and time\footnote{Apart from the possible correlations in $d>1$
between the different components of the displacement $\bu(\bx)$ at a
single point $\bx$. However, if we consider the case that $p(\bu)$
depends only on $u=|\bu|$ (i.e., the displacement field is isotropic),
perpendicular displacements are uncorrelated.}. 
Moreover $f(\bu,\bv;\bx)$ is a discontinuous function at $\bx=0$, 
that is
$G_{\mu\nu}(\bx)=0$ for $\bx\neq 0$, whereas 
$G_{\mu\nu}(0) \ne 0$ and in particular 
$G_{\mu\nu}(0)=\delta_{\mu\nu}\overline{u^2}/d>0$ when assuming 
that $p(\bu)=p(u)$. Then, we can simply show that 
$\hat\phi(\bk;\bx)=|\hat p(\bk)|^2$ 
where $\hat p(\bk)={\cal
F}[p(\bu)]$, and, because of the discontinuity in $\bx=0$, $\int
d^dq\,\tilde\phi(\bk;\bq)=|\hat p(\bk)|^2\ne \hat\phi(\bk;0)=1$. 
This
implies that Eq.~(\ref{eq3}) reads
\be
S_f(\bk)=1+|\hat p(\bk)|^2[S_{in}(\bk)-1]\;.
\label{eq3b}
\ee
Note that this relation is local in $\bk$, i.e., each $\bk-$mode of 
the particle density evolves independently one of each other.
\item  
$\bu(\bx)$ is a real correlated and continuous stochastic process.
In this case \cite{gnedenko}, it is well known that
$G_{\mu\nu}(\bx)$ is a continuous function of $\bx$ and 
in the limit when $\bx\rightarrow 0$, 
it approaches with continuity its value $G_{\mu\nu}(0)$ 
\cite{gnedenko}. 
In other words, $f(\bu,\bv;\bx)$ is continuous in $\bx$ and
$$
\lim_{\bx\rightarrow 0}f(\bu,\bv;\bx)=\delta(\bu-\bv)p(\bu)\,.
$$
this implies that
$\int d^dq\,\tilde\phi(\bk;\bq)= \hat\phi(\bk;0)=1$  
\cite{displa},  and therefore
Eq.~(\ref{eq3}) becomes
\be
S_f(\bk)= n_0\int d^dx\, e^{-i\bk\cdot\bx}\hat\phi(\bk;\bx)[1+\xi_{in}(\bx)]
-(2\pi)^d n_0\delta(\bk)\,.
\label{eq3c}
\ee
Unlike Eq.~(\ref{eq3b}), this equation is nonlocal in $\bk$
due to the presence of displacement-displacement spatial 
correlations which couples different modes of the particle density
before and after the application of the displacement field.
\end{enumerate}

We can now study the spatial diffusion of a particle distribution
in which at each time steps the particles move under one of the 
stochastic displacement field just described above. 
We will assume for simplicity, that the time can be discretized in
time-steps of size $\Delta t$ at which the particle distribution 
is displaced by a realization of the field $\bu(\bx,t)$; 
moreover different time-steps are supposed to be statistically independent.

\section{Spatially uncorrelated displacements}
In order to illustrate the general formalism to derive a continuous
time equation from the discrete one (single step), it is instructive to
consider the simplest case of spatially and temporally uncorrelated
displacement field. As we show below, it corresponds to the
homogeneous diffusion equations, either standard or fractional
depending whether the variance of displacements is finite or
infinite. This discussion is useful in view of the more interesting
case of random walks generated by spatially
correlated displacements.  As clarified above, all the statistics of
the field is contained in the one-displacement PDF $p(\bu)$. Let us
firstly consider the statistically isotropic case for $\bu$:
$p(\bu)=p(u)$, implying that $\hat p(\bk)=\hat p(k)$. We have to
distinguish the two cases of finite and infinite variance
$\overline{u^2}$. In the former, each particle performs an ordinary
$d-$dimensional random walk and the paths of different particles are
independent one of each other, while in the latter, each particle undergoes
a $d-$dimensional Levy walk independently of the others.

\subsection{The continuous time limit}

For statistical isotropic displacements, we can expand  
at small $k$ as
\be 
\hat
p(k)=1-Bk^\alpha+o(k^\alpha)\,,
\label{uncorr1}
\ee 
where $\alpha=2$ and $B =\overline{u^2}/2d$ when 
$\overline{u^2}$ is finite, while $0<\alpha <2$ when 
$\overline{u^2}$ diverges and 
$p(u)\simeq Au^{-(\alpha+d)}$ at large $u$,  
with $B$ proportional to the amplitude of the tails $A>0$. 
In the first case each
particle undergoes an independent standard random walk 
while, in the second, an independent Levy flight \cite{Levyflights}.
A well defined diffusional continuous time
limit is attained by requiring that $2B=D\Delta t$
where $D>0$ is a constant independent of the time step $\Delta t$.  
Therefore in the limit $\Delta t\to 0$, 
the substitution of Eq.~(\ref{uncorr1}) into 
Eq.~(\ref{eq3b}) leads, for 
$k^\alpha\ll 2/(D\Delta t)$, to
\be 
\partial_t
S(\bk,t)=Dk^{\alpha}\left[1-S(\bk,t)\right]\,,
\label{uncorr4}
\ee
whose solution is
\be
S(\bk,t)=\left[S_{in}(\bk)-1\right]\exp(-D k^{\alpha}t)+1\,.
\label{uncorr4b}
\ee 
The approach of $S(\bk,t)$ to homogeneous
Poisson power spectrum $S(\bk)=1$
is exponentially fast, and each $\bk-$mode of the 
two-point correlation relaxes with a rate  
$Dk^{\alpha}$  ($0<\alpha\le 2$).
We see immediately that the smaller $\alpha$ the faster the 
approach to the completely uncorrelated stationary state.
In other words inhomogeneities diffuse ($\alpha=2$) or super-diffuse
($0<\alpha<2$) until reaching
the uniform stationary state $\Gamma_s(\bx)= n_0$.
Equation (\ref{uncorr4}) can be recast in a more familiar form
by rewriting Eq.~(\ref{uncorr4b}) for $\hat h(\bk,t)$: 
\be 
\partial_t \hat
h(\bk,t)=-Dk^{\alpha}\hat h(\bk,t)\,.
\label{uncorr4c}
\ee
and then taking the inverse FT.
For simplicity, let us consider a one dimensional system.
By taking the FT of Eq.~(\ref{uncorr4c}) we can write
\be 
\partial_t h(x,t)=D\partial^{\alpha}_x h(x,t)\,,
\label{uncorr4d}
\ee
where $\partial^{\alpha}_x$ is for $\alpha=2$ the usual Laplacian in $d=1$,
giving the celebrated standard diffusion equation,
while $\partial^{\alpha}_x$ for $0<\alpha<2$ is the fractional derivative 
\cite{FracRW1} of
order $\alpha$ in $x$.  Equation (\ref{uncorr4d}) is for $0<\alpha<2$
the well known 
equation of spatially fractional diffusion \cite{FracRW1,FracRW2} 
and describes how the connected two-point correlations super-diffuse 
in space toward an uncorrelated Poisson steady
state. The equation~(\ref{uncorr4d}) takes a more familiar 
form when written for the average conditional
density $\Gamma(x,t)$: 
\be 
\partial_t
\Gamma(x,t)=D\partial^{\alpha}_x \Gamma(x,t)\,,
\label{uncorr4e}
\ee 
which is the fractional diffusion equation for the density of
particles seen in average by a generic particle of the system.  It
belongs to a larger family of fractional diffusion equations whose
general form is
\be
\partial_t^\beta
\Gamma(x,t)=D\partial^{\alpha}_x \Gamma(x,t)
\label{frac-dif}
\ee with $0<\beta\le 1$ and $0<\alpha\le 2$.  This kind of equation is
encountered in the description of key aspects of anomalous transport
as for instance that occurring on disordered peculiar structures 
(fractal supports) or due to spatial non locality \cite{FracRW1,FracRW3}.
More precisely the fractional diffusion equation (\ref{frac-dif}) can
be obtained in the context of continuous time random walks (CTRW)
\cite{CTRW} under the assumption that the joint PDF $\psi({\bu},t)$
to make a step of size ${\bu}$ in the time interval $[t,t+dt)$
factorizes as $\phi(t)p({\bu})$. In order to obtain $\beta=1$
it is necessary that the mean value of $\phi(t)$ is finite, and to
obtain $\alpha=2$ we need a finite variance for $p(\bu)$.
Otherwise we have respectively $\beta<1$ and $\alpha<2$ depending on
the power law tails of the two functions $\phi(t)$ and $p(\bu)$.  
Indeed we have obtained Eq.~(\ref{uncorr4e}) via the choice
$\phi(t)=\delta(t-\Delta t)$ (with $\Delta t \to 0^+$) which has finite
mean value equal to $\Delta t$ itself.
            
\section{Spatially correlated displacements} \label{corr}
We now turn to the case where the stochastic displacements, acting on the
particles at each time-step, are generated through the realizations of
a continuous and spatially stationary correlated stochastic field
$\bu(\bx)$ defined by a probability density functional ${\cal
P}[\bu(\bx)]$.  For instance, in $d=1$ we can consider a Gaussian
stochastic field defined by the probability density functional
$$
{\cal P}[u(x)]\sim \exp\left[-\int\int
dx\,dy\, u(x)K(x-y)u(y)\right]
$$ 
with $K(s)$ being the positive definite and continuous correlation
kernel.  Since we assume again no time correlation between successive
realizations of the field $\bu(\bx)$, at each time-step the power spectrum 
of the particle distribution evolves according to Eq.~(\ref{eq3c}).  In
principle, a time dependent PDF ${\cal P}[\bu(\bx);t]$ can be also
considered, referring to a displacement field whose correlation
properties depend on time.  Here we limit the discussion to
time independent functionals ${\cal P}$, focusing on the case of a
$d-$dimensional Gaussian displacement field with finite variance
$\overline{u^2}=\sum_{\mu=1}^{d}G_{\mu\mu}({\bf 0})<+\infty$, where
the displacement-displacement correlation matrix $G_{\mu\nu}(\bx)$ has 
been defined in Sec.~\ref{sec1}. However, as shown explicitly below, the
validity of the evolution equations we derive
is not restricted to this case. If the displacement field is
Gaussian its characteristic function reads \cite{displa}: 
\be
\hat\phi(\bk;\bx)=\exp\left\{-\sum_{\mu,\nu}^{1,d}k_\mu k_\nu
[G_{\mu\nu}({\bf 0})-G_{\mu\nu}({\bf x})]\right\}\,,
\label{eq4}
\ee
where $k_\mu$ indicates the $\mu^{th}$ component of $\bk$.

In analogy with the simpler case of uncorrelated displacements,
we have to assume 
$$
G_{\mu\nu}(\bx)=\Delta t c_{\mu\nu}(\bx)
$$ for the existence of a smooth time limit, with $c_{\mu\nu}(\bx)$
independent on $\Delta t$ in order to obtain the correct diffusional
processes in the limit $\Delta t\to 0$.  In fact the above condition
implies that the variance of each component $\mu$ of the field
satisfies $G_{\mu\mu}(0)=\overline{|u_{\mu}(\bx,t)|^2}\sim \Delta t$
at any time $t$.  Therefore, if $\bu(\bx,n\Delta t)$ is the
displacement in the point $\bx$ at time $t=n\Delta t$, we can write:
\bea 
&&\overline{\bu(\bx,n\Delta t)}=0\\ 
&&\overline{u_\mu(\bx,n\Delta
t)u_\nu(\bx',n'\Delta t)}= \Delta t
c_{\mu\nu}(\bx-\bx')\delta_{nn'}\,,
\label{eq5}
\eea thus, in the limit $\Delta t\rightarrow 0$ the field
$\eta(\bx,t)= \bu(\bx,t)/\sqrt{\Delta t}$ becomes a spatially
correlated and temporally delta-correlated noise for the motion of
particles.  In a hydrodynamic analogy, in which the displacement field
is interpreted as a turbulent main flow advecting the particles of a
passive pollutant, the quantities
$d_{\mu\nu}(\bx)=[c_{\mu\nu}(0)-c_{\mu\nu}(\bx)]$ are usually called
\cite{verg-pre96} the {\em structure functions} of the flow.

In the limit of small $\Delta t$, we can expand Eq.~(\ref{eq4})
\be 
\hat\phi(\bk;\bx) = 1 - \Delta t\sum_{\mu,\nu}^{1,d}k_\mu k_\nu
d_{\mu\nu}({\bx})+o(\Delta t)\,.
\label{eq4b}
\ee
When plugged into Eq.~(\ref{eq3c}) this gives
\bea
\label{eq4c}
S(\bk,t+\Delta t) &=& S(\bk,t) + \Delta t\sum_{\mu,\nu}^{1,d}k_\mu k_\nu
\left[n_0\hat c_{\mu\nu}(\bk)- c_{\mu\nu}(0)S(\bk,t)\right] + \\
&+&\Delta t\sum_{\mu,\nu}^{1,d}
k_\mu k_\nu \int \frac{d^dq}{(2\pi)^d}S(\bk-\bq,t)\hat c_{\mu\nu}(\bq)
+o(\Delta t),
\nonumber
\eea
where $\hat c_{\mu\nu}(\bk)={\cal F}[c_{\mu\nu}(\bx)]$ is the re-normalized
power spectrum tensor of the instantaneous displacement field.  
The limit
$\Delta t\rightarrow 0$ on Equation~(\ref{eq4c}) yields 
\be
\partial_t S(\bk,t)=\sum_{\mu,\nu}^{1,d} k_{\mu} k_{\nu}\left[
 n_0\hat c_{\mu\nu}(\bk) -c_{\mu\nu}(0)S(\bk,t)+
\int \frac{d^dq}{(2\pi)^d}\hat S(\bk-\bq,t)\hat c_{\mu\nu}(\bq)\right],
\label{eq15a}
\ee
which in terms of FT of the covariance function takes the same forms 
\be
\partial_t \hat h(\bk,t)=\sum_{\mu,\nu}^{1,d} k_{\mu} k_{\nu}\left[\hat 
c_{\mu\nu}(\bk) -c_{\mu\nu}(0)\hat h(\bk,t)+
\int \frac{d^dq}{(2\pi)^d}\hat h(\bk-\bq,t)\hat c_{\mu\nu}(\bq)\right].
\label{eq15}
\ee
and under the inverse FT it becomes
\be
\partial_t h(\bx,t)=\sum_{\mu,\nu}^{1,d}\partial^2_{\mu\nu} 
\left[d_{\mu\nu}(\bx)
h(\bx,t)-c_{\mu\nu}(\bx)\right]\,.
\label{eq16}
\ee
Consequently, for $\Gamma(\bx)= n_0[1+h(\bx)]$ we obtain
\be
\partial_t \Gamma(\bx,t)=\sum_{\mu,\nu}^{1,d}\partial^2_{\mu\nu} \left[ 
d_{\mu\nu}(\bx)\Gamma(\bx,t)\right]\,.
\label{eq17}
\ee 
It is noteworthy that if $P(\bx,t|\bx_0,t_0)$ is the PDF of
the separation $\bx$ at time $t$ between an arbitrary pair of particles of the 
distribution, given their initial distance $\bx_0$ at time $t_0$,
we can write
\[
\Gamma(\bx,t)=\int d^dx_0\,P(\bx,t|\bx_0,t_0)\Gamma(\bx_0,t_0)\,.
\]
It is straightforward to verify that the transition probability 
$P(\bx,t|\bx_0,t_0)$ also satisfies Eq.~(\ref{eq17}) and  
for this reason $P(\bx,t|\bx_0,t_0)$ is called the {\em propagator} of 
the diffusion operator defined by Eq.~(\ref{eq17}).
On the other hand  Eq.~(\ref{eq17}) is the Fokker-Planck (FP) equation
associated to the stochastic Langevin equation (LE) in \^Ito representation 
\cite{Gardiner} 
\be
\dot\bx(t)=\bw(t)
\label{eq-lang}
\ee
describing the time evolution of the two particle separation $\bx(t)$,
where $\bw(t)$ is a Gaussian noise with the following one and two-time 
correlation properties:
\be
\left\{
\begin{array} {ll}
\overline{w_\mu(t)}=0& \\
 &\mu,\nu=1,...,d\\
\overline{w_\mu(t)w_\nu(t')}=2d_{\mu\nu}(\bx)\delta(t-t')&
\end{array}
\right.
\label{eq-lang2}
\ee
Therefore Eqs.~(\ref{eq17}) and (\ref{eq-lang}) are equivalent.

In the hyper-isotropic condition, corresponding to the choice
$d_{\mu\nu}(\bx)=\delta_{\mu\nu} d(x)$ [i.e.,
$c_{\mu\nu}(\bx)=\delta_{\mu\nu} c(x)$], Eq.~(\ref{eq17}) can be
rewritten as 
\be
\partial_t\Gamma(\bx,t)=\nabla^2[d(x)\Gamma(\bx,t)]\,.
\label{eq-isotropy}
\ee 
Note that all these results are
not restricted to the case of Gaussian displacement fields. In
fact, as shown in \cite{displa}, if $G_{\mu\nu}(\bx)=
\overline{u_{\mu}(\bx_0)u_{\nu}(\bx_0+\bx)}$ is finite for $x=0$, the
small $k$ expansion
\[\hat\phi(\bk;\bx)=1-\sum_{\mu,\nu}^{1,d}k_\mu k_\nu
[G_{\mu\nu}({\bf 0})-G_{\mu\nu}({\bx})]+o(k^2)\] 
is always valid.

It is easy to prove, but important to note, that when the displacement 
field $\bu(\bx,t)$ can be decomposed into two independent components: 
a spatially correlated field $\bu_1(\bx,t)$ characterized by Eq.~(\ref{eq5}) 
and a spatially uncorrelated 
and statistically isotropic field $\bu_2(\bx,t)$ of variance 
$\overline{u_2^2}=d\times D\Delta t$, 
Eq.~(\ref{eq17}) takes the form: 
\be
\partial_t \Gamma(\bx,t) = D\nabla^2 \Gamma(\bx,t)+
\sum_{\mu,\nu}^{1,d}\partial^2_{\mu\nu} \left[ 
d_{\mu\nu}(\bx)\Gamma(\bx,t)\right]\,.
\label{eq17b}
\ee 
That is, the motion of the set of particles results in the
superposition of a standard diffusion, first term in
Eq.~(\ref{eq17b}), with a spatially correlated diffusion, second term.
Equations of the type (\ref{eq17b}) are the generalized diffusion
equations which are usually encountered in the context of turbulent
transport of passive scalars (i.e., pollutant) \cite{Scalars}, where $
n(\bx,t)$ is the density of the passive particles advected by the
velocity field $\bv(\bx,t)=\bu(\bx,t)/\Delta t$ of the synthetic 
turbulent
flow. It is important to note that for $x \to 0$, the term $ D\nabla^2
\Gamma(\bx,t)$ of Eq.~(\ref{eq17b}) dominates because
$d_{\mu\nu}(0)=0$, therefore the small scale motion occurs via
standard diffusion. The interesting case is obtained when $D \to 0$,
and consequently $\bx = 0$ becomes a singularity of
Eq.~(\ref{eq17b}) \cite{Gardiner}. 
In particular in turbulence a complete solution has
been given for the scale free case in which 
\be 
d_{\mu\nu}(\bx) = a
x^{\xi} \delta_{\mu\nu} + b x^{\xi-2} x_{\mu} x_{\nu}\,,
\label{eq:d_krai}
\ee
with $0<\xi\le 2$, $a$ and $b$ constants such that the tensor 
$c_{\mu\nu}(\bx)$ has a positive definite FT. 
The one dimensional case 
is recovered by putting $d=1$ and $b=0$. The class of models defined by 
the structure tensor (\ref{eq:d_krai}) is referred to as 
{\em generalized Kraichnan ensemble}  \cite{kraich,Scalars}
and a complete classification of their solutions, in terms of the singular 
behavior around $\bx =0$, has been given in Ref.~\cite{verg-gaw,gaw}.  
For $0<\xi<2$, it consists in three possible behaviors: 
\begin{enumerate}
\item For $(b/a)>(d-2)$ and at the same time $(\xi-1)(b/a)\ge (d-\xi)$,
the only possible solution of Eq.~(\ref{eq17}) is such that different particles
coalesce in finite time and no stationary state exists. 
In practice pair of particles collide in finite time with vanishing 
relative velocity and therefore remains attached for the rest of the dynamics. 
In this case $\Gamma(\bx)\simeq \alpha(t)\delta(\bx)+
\beta(t)x^{2-\xi-d}$ with
a time increasing coefficient $\alpha(t)$, which signals the coalescence
phenomenon;
\item For $(\xi-1)(b/a)<(d-\xi)$ and at the same time $(b/a)\le (d-2)$ 
the only possible solution is such that the probability of finding 
more than one particle in a single spatial point is zero at all time. 
In this phase particles form only clusters and the diffusion of 
particles converges to a stationary state in which 
$\Gamma(\bx)\sim x^{-\gamma}$ where $\gamma=\xi+b(d-1)/(a+b)$;
\item Finally for $(b/a)>(d-2)$ and $(\xi-1)(b/a)<(d-\xi)$ 
particles can collide at finite time, but with non-zero relative velocity. 
This obliges to fix a boundary condition at $\bx=0$ for the diffusion 
equation. For an {\em absorbing} boundary condition one has an effective
behavior as in case (i) above. Instead for a {\em reflecting} boundary 
condition the effective behavior is similar to case (ii) above.
Adopting a mixed boundary condition a composition of the two above 
behaviors appears (called ``sticky'' phase in \cite{gaw}).
\end{enumerate}
The method used to get this classification is quite complex and
consists in the theory of boundary conditions of elliptic operators.
In the rest of the paper we do not enter the details of this rigorous
analysis, but limit our study to the one dimensional case for a
generic choice of the displacement-displacement correlation function
through computer simulations and simple theoretical arguments.

\subsection{The one dimensional spatially correlated diffusion}

In $d=1$ Eq.~(\ref{eq17}) becomes 
\be
\partial_t\Gamma(x,t)=\partial_x^2[d(x)\Gamma(x,t)]\,,
\label{eq:1d}
\ee where $d(x)=[c(0)-c(x)]$ is the structure function of the
stochastic velocity field (i.e., displacements) with two-point
correlation function $c(x)$. It is important to note that
the general properties of any correlation function $c(x)$
constraints $d(x)$ to have the small $x$ behavior $d(x)\simeq a x^\xi$, 
with $a>0$ and $0<\xi\le 2$. 
In this paper we do not treat the ``smooth'' case $\xi=2$ as it 
has been solved rigorously elsewhere \cite{coha} and focus our analysis on 
the ``rough'' stochastic velocity fields for $0<\xi<2$.
The aforementioned classification of the solutions of the Kraichnan
ensemble immediately implies that for $b=0$ and $d=1$ in
Eq.~(\ref{eq:d_krai}), only the first and the third case are possible
in one dimension around the singularity $x=0$. In particular we have
the first behavior for $1\le \xi< 2$, and the third for $0<\xi<1$. For $\xi=2$
(when the velocity field is smooth at small scales) one can see that
particles coalesce but in an infinite mean time \cite{coha}.

First of all we note that, if it exists, the only possible stationary
solution of Eq.~(\ref{eq:1d}) is 
\be
\Gamma_s(x)= n_0\frac{c(0)}{d(x)}\,
\label{eq:1d-st}
\ee 
$\Gamma_s(x)$, due to its definition of average conditional
density, must converge to $ n_0$ for $|x|\to\infty$ where
density-density correlations has to disappear.  
In order to decide around the acceptability or not of this stationary 
solution, it is necessary to study its small scale behavior.  
We know that by
definition $\Gamma_s(x)$ is acceptable only if it is integrable at
small $x$.  We see immediately that in $d=1$ it happens only for
$\xi<1$. Thus for $\xi\ge 1$, Eq.~(\ref{eq:1d}) admits no
stationary solution in agreement with the aforementioned
classification of solutions of the generalized Kraichnan's ensemble
\cite{verg-gaw,gaw}. 
In fact Eq.~(\ref{eq:1d-st}) corresponds to the
stationary state which the family of solutions with simple
clustering and no coalescence converge to,
obtained by imposing a reflecting boundary condition at the
singularity $x=0$.  
This solution is analogous to the Poisson
stationary correlation function $\Gamma_s(x)= n_0$ for the ordinary
diffusion equation to which corresponds the well known
propagator
\be
P(x,t|0,t)=\frac{1}{\sqrt{4\pi Dt}}\exp\left(-\frac{x^2}{4Dt}\right)\,,
\label{gauss}
\ee
satisfying the reflecting condition at $x=0$ and which
shows how two particles at initial vanishing distance spread when a reflecting
condition is imposed about their collisions. 
An analogous propagator can be found also in relation to the stationary
state (\ref{eq:1d-st}) of our more complex diffusion Eq.~(\ref{eq:1d}).
It can be found by looking for a scaling solution of the form
$P(x,t|0,t)=t^{-\beta} f(x/t^\beta)$. Plugging this scaling form into
Eq.~(\ref{eq:1d}), one finds
\be
P(x,t|0,t)=\frac{C}{t^{(1-\xi)/(2-\xi)}}x^{-\xi}
\exp\left(-\frac{x^{2-\xi}}{a(2-\xi)^2t}\right)
\label{eq:1d-prop}
\ee 
where $C>0$ is the normalization constant.  Also
Eq.~(\ref{eq:1d-prop}) describes the spreading of a pair of particles
at initial vanishing distance with reflecting condition at $x=0$. In
fact $\partial_x [x^\xi P(x,t|0,t)]|_{x=0}=0$.  Moreover it becomes
the ordinary diffusing Gaussian for $\xi\to 0$. Let us now analyze in
detail Eq.~(\ref{eq:1d-st}).  First of all we note that for all $x$ the
covariance function $h(x)=[\Gamma(x)-n_0]/n_0$ has the same
sign of the displacement correlation function $c(x)$, i.e., those 
scales at which displacements are positively (negatively) correlated
asymptotically become also positively (negatively) correlated scales
for the density of particles. Moreover $\Gamma_s(x)$ for large $x$
approaches the average density $ n_0$ in the following way
\[\Gamma_s(x)\simeq  n_0\left[1+\frac{c(x)}{c(0)}\right]\,,\]
i.e.,  $h(x)$ is
\[h(x)\simeq \frac{c(x)}{c(0)}\,.\]
In other words the iterated displacement field injects exactly its
large scale correlations in the particle system. This is interesting
because, as one can check by expanding Eq.~(\ref{eq3c}) at small $k$
\cite{displa}, the displacement 
field injects, in a single step, only a large scale contribution to 
the power spectrum $S(k)$ [or $\hat h(k)$] of order $k^2\hat c(k)\Delta t$ 
which vanishes $k^2$ times faster than $\hat c(k)$.
Finally, at small $x$ we have $\Gamma_s(x)\sim x^{-\xi}$, meaning that 
particles forms, at small scales, clusters with fractal dimension 
$D=(1-\xi)$.

We now turn to the problem of what happens for $\xi\ge 1$ and how the
other ``non-reflecting'' solutions for $\xi<1$ behave. This is a more
difficult task as the singularity at $x=0$ of Eq.~(\ref{eq:1d})
generates a coalescence dynamics which can not be described through
only smooth functions.  In order to study this case, we adopt a sort
of mean field approximation in an appropriately transformed LE 
(in \^Ito representation) for the separation $x$ between a
pair of particles associated to the FP Eq.~(\ref{eq:1d}).  We suppose
that at initial time $t=0$ we have $x_0\equiv x(0)>0$.  At 
sufficiently small $x$ Eq.~(\ref{eq-lang}), for our one dimensional
case, can be written as 
\be 
\dot x(t)=A[x(t)]^{\xi/2}\eta(t)\,,
\label{eq-lang:1d}
\ee
where $A=\sqrt{2a}>0$ and  $\eta(t)$ is a white Gaussian noise such that
$\overline{\eta(t)}=0$ and $\overline{\eta(t)\eta(t')}=\delta(t-t')$.
Let us now apply the following change of variables to Eq.~(\ref{eq-lang:1d}):  
\be
y=\frac{2}{A}\frac{x^{1-\xi/2}}{2-\xi}\,,
\label{eq:y-x}
\ee
i.e.,
\be
x=\left[{A\over 2}(2-\xi)y\right]^{2/(2-\xi)}\,.
\label{eq:x-y}
\ee
By using the rule of change of variables in the \^Ito representation
\cite{Gardiner}, we get the LE for $y(t)$
\[\dot y(t)=-\frac{C}{y(t)}+\eta(t)\,,\]
where $C=\xi/(4-2\xi)$.
If we suppose that the initial value $y(0)>0$ is sufficiently small, 
the noise $\eta(t)$ can be neglected obtaining 
\[\dot y(t)=-\frac{C}{y(t)}\]
whose solution is
\be
y(t)=\sqrt{y^2(0)-\frac{\xi}{2-\xi}t}\,,
\label{eq:y}
\ee
where $y(0)=\frac{2}{A}\frac{x_0^{1-\xi/2}}{2-\xi}$ from 
Eq.~(\ref{eq:y-x}).
This equation implies that at $t^*=(2-\xi)y^2(0)/\xi$ we get
$y(t^*)=0$ and therefore $x(t^*)=0$, i.e. the two particles collide.
Moreover the velocity $\dot x(t)$, in the approximation where we 
neglect the noise, is 
\[\dot x(t)={dx\over dy}\cdot{dy\over dt}\sim -\left[y^2(0)-
{\xi\over 2-\xi}t\right]^{(\xi-1)/(2-\xi)}\,.\] 
Therefore $\dot x(t^*)=0$ for $1\le \xi<2$, while $\dot
x(t^*)\to -\infty$ for $\xi<1$.  Consequently, for $1\le \xi<2$,
independently of the boundary condition at $x=0$, 
once $x(t)$ vanishes it keeps this value forever, i.e.,
for $1\le \xi<2$ this coalescing solution is the only possible solution
as predicted by the general classification given above for the
Kraichnan ensemble. Instead
for $\xi<1$, since the velocity $\dot x(t)$ is non-zero, 
when $x(t)=0$, it is necessary
to choose by hand the boundary condition to fix the kind of
solution. For reflecting boundary condition we have the above
presented solution converging to the stationary state $\Gamma_s(x)$
given in Eq.~(\ref{eq:1d-st}), while for absorbing boundary condition
no stationary solution is reached and we have a behavior similar to
that for $\xi\ge 1$.
We now derive the small $x$ behavior of the propagator 
$P(x,t|x_0,0)$ with sufficiently small $x_0>0$ from Eq.~(\ref{eq:y}) 
valid for both cases $0<\xi<1$ with absorbing boundary condition and 
for $1<\xi<2$.
Equation (\ref{eq:y}) says that at time $t$ all pair of particles with
$y^2(0)\le \frac{\xi}{2-\xi}t$, i.e., with initial separation
$x_0\le\left[\frac{a\xi(2-\xi)}{2}t\right]^{1/(2-\xi)}\equiv x_{max}(t)$ 
have already collided at time $t$. 
For $1\le\xi<2$ such pairs of particles always coalesce, as 
$\dot x(t)=0$ when $x(t)=0$. Instead for $0<\xi<1$ 
they coalesce only if we impose the absorbing condition
(i.e. completely inelastic collision) at $x=0$. In both cases we can say
that $P(x,t|x_0,0)$ has developed a singular contribution
$m(t)\delta(x)$ at $x=0$ where 
\[m(t)=\int_0^{x_{max}(t)} dx_0\, p(x_0)\,,\]
where $p(x_0)$ is the PDF of the initial pair distance $x_0$.
If $\Gamma_s(x,0)= n_0$ (i.e. initial particle distribution is a homogeneous
random Poisson one) for small $x_0$ (i.e. $x_0\ll 1/n_0$)
we have $p(x_0)\simeq n_0$, from 
which we obtain $m(t)\simeq n_0x_{max}(t)\sim t^{1/(2-\xi)}$.
On the other side for all pairs with $x_0>x_{max}(t)$ we can 
develop the following argument. Since in the present approximation 
the dynamics is deterministic one can derive $P(x,t|x_0,0)$
at $x>0$ directly by a simple change of variable in $p(x_0)$.
More precisely let us call $x_t=x(t)$ and $P(x,t|x_0,0)=p_t(x_t)$;
we can write by conservation of probability:
\be
p_t(x_t)=p(x_0){dx_0\over dx_t}
\label{eq:p-x-t}
\ee
and the relation between $x_t$ and $x_0$ is given by Eqs.~(\ref{eq:y}),
(\ref{eq:y-x}) and (\ref{eq:x-y}).
By considering again $p(x_0)\simeq n_0$, it is simple to derive
for small $x>0$ that
\[
p_t(x)\sim t^{(\xi-1)/(2-\xi)} x^{1-\xi}\,.
\]
Note that $\lim_{x\to 0}p_t(x)=0$ for $\xi<1$ as it has to be for
an absorbing condition at $x=0$.
Finally at small $x$ and sufficiently small $t$, if 
$\Gamma(x\to 0,0) = n_0$, the expression   
\be
\Gamma(x,t)\sim P(x,t|x_0,t)=bt^{1/(2-\xi)}\delta(x)+
ct^{(\xi-1)/(2-\xi)} x^{1-\xi}\,,
\label{eq:gamma-x-t}
\ee
represents the solution for $1<\xi<2$, while it is a
solution for the case $0<\xi<1$, only when considering absorbing 
conditions at $x=0$". With $b$ and $c$ positive constants.
This completely agrees with what found in \cite{verg-gaw,gaw}. 

\subsection{Numerical results}
In the light of the previous one dimensional analysis, we discuss an
important example in $d=1$ where $u(x)$ is a spatially correlated
Gaussian displacement field with short range correlation function
\be 
c(x)=c(0) \exp\left\{
-\left|\frac{x}{x_0}\right|^\xi\right\}, 
\quad\quad 0<\xi\le2\,.
\label{eq:corrfield}
\ee where $x_0$ determines the correlation length.  We recall that
according to the Wiener-Khinchin theorem \cite{gnedenko}, a given
$c(x)$, such that $c(0)>0$, is a well defined correlation function of
a continuous stochastic field if and only if $c(x)$ is continuous for 
all $x$, $\hat c (k)\ge 0$ for any $k$, and finally
$\int_{-\infty}^{+\infty} dk\,\hat c(k)<+\infty$.  All these conditions
are satisfied by Eq.~(\ref{eq:corrfield}).  Varying $\xi$ between $0$
and $2$ allows us to study numerically all the phases above
described theoretically:\newline 
(i) for $0<\xi<1$, in case of reflecting boundary conditions on
particle-particle collisions, the particle diffusion admits the
stationary solution Eq.~(\ref{eq:1d-st}) implying a small $x$ behavior
\be 
\Gamma_s(x)\simeq n_0\left|\frac{x}{x_0}\right|^{-\xi}\,,
\label{ex2}
\ee 
which describes a fractal distribution with dimension $D_f=1-\xi$
for $x<x_0$, and exhibits a cross over to a uniform particle
distribution with average density $ n_0$ for $x>x_0$. For $x\gg x_0$
instead correlations decay exponentially fast as $h_s(x) \simeq
\exp\left(-\left|\frac{x}{x_0}\right|^\xi\right)$.; \newline 
(ii) {for $1\le\xi<2$, the system admits no stationary behavior and the unique
solution corresponds to particle coalescence} as described by
Eq.~(\ref{eq:gamma-x-t}).
\begin{figure}
\includegraphics[clip=true,keepaspectratio,width=8.cm]{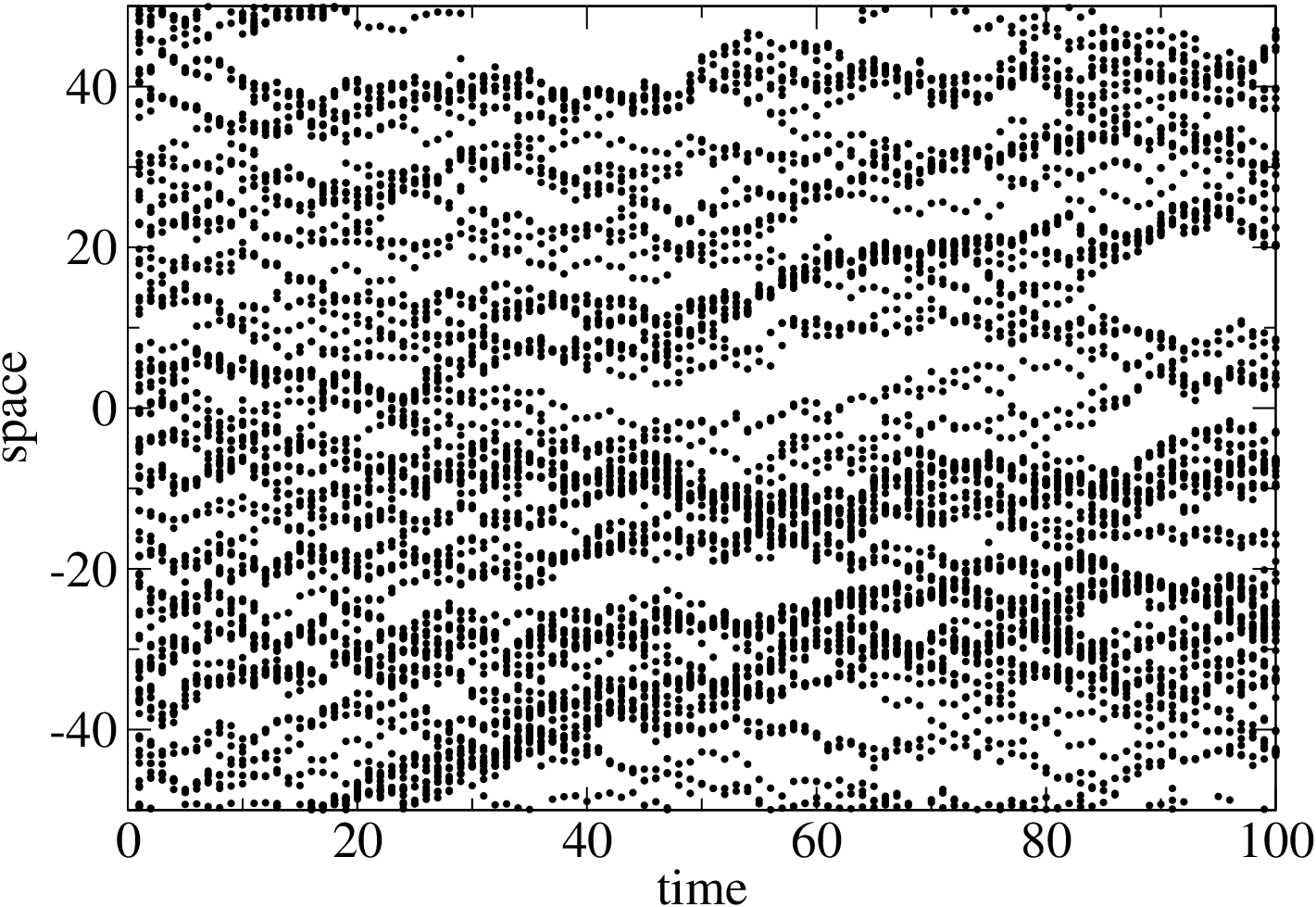}\hspace{0.cm}
\includegraphics[clip=true,keepaspectratio,width=8.cm]{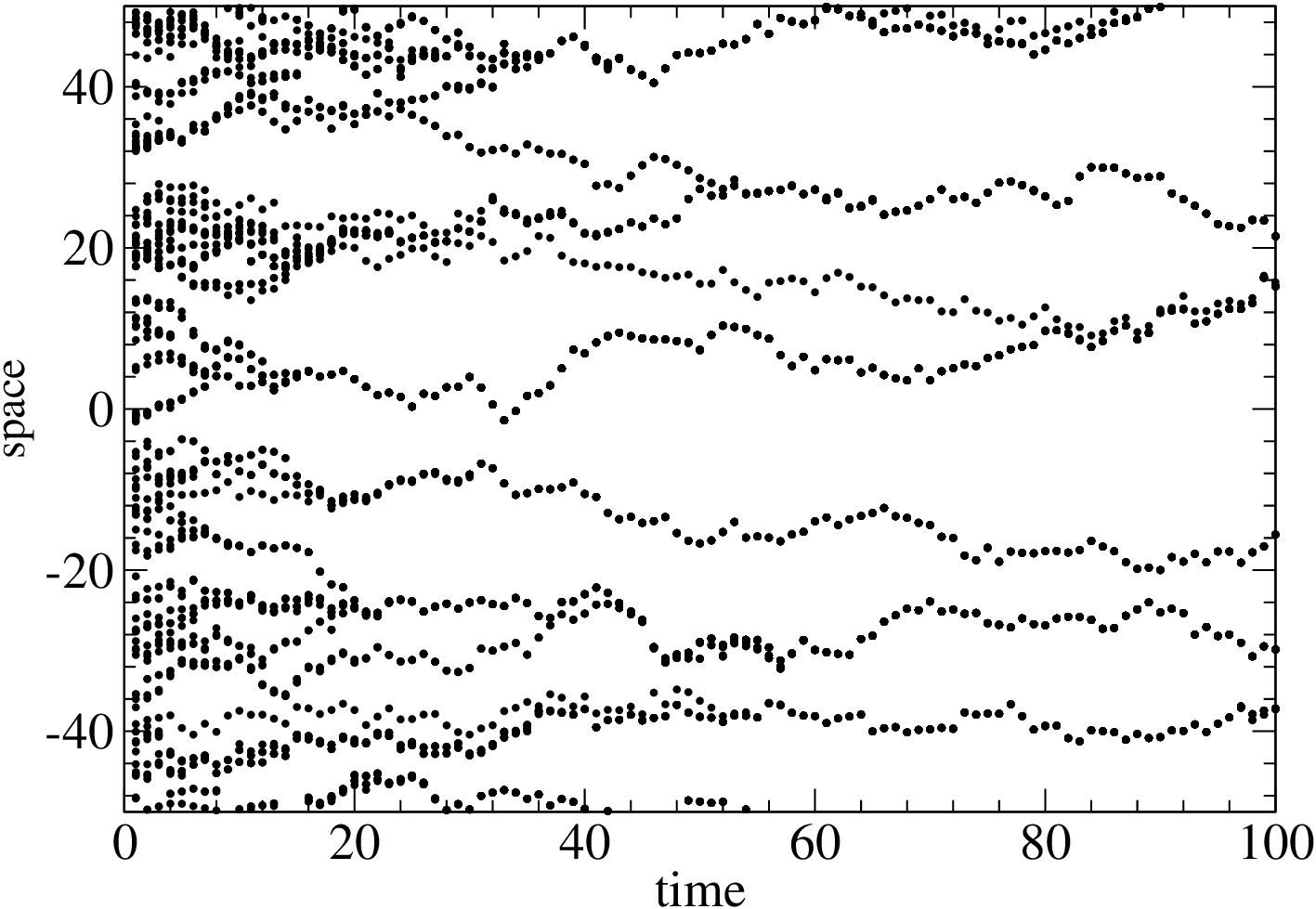}
\caption{Detail of the  space-time pattern generated by the trajectories of $N=512$
random walkers undergoing the evolution Eq.~(\ref{eq:move_h}) with 
time step $h=10^{-4}$ and displacements that are not 
time-correlated but have a spatial correlation (\ref{eq:corrfield}),
with $x_0=3$. Left panel corresponds to $\xi=0.5$ showing 
simple clustering and right refers to $\xi=1.5$, for which 
a coalescence regime occurs.}
\label{fig:evol}
\end{figure}

We run computer simulations to check these behaviors of particle
distributions under the effects of a repeated application of the
displacement field with correlation (\ref{eq:corrfield}). The
numerical implementation of the dynamics requires the generation of an
array of correlated Gaussian random variables $\{u_i\}$ 
with the prescribed correlator $\langle u_i u_j \rangle\sim c(x_i-x_j)$ 
where $c(x)$ is given by Eq.~(\ref{eq:corrfield}) and $x_i$ is the position
of the $i^{th}$ particle ($i=1,...,N$). 
We used two methods:
\begin{enumerate}
\item The first one
considers the Cholesky decomposition method \cite{Cholesky} for the
covariance matrix. According to this algorithm, an array
$\{u_1,....,u_N\}$ of $N$ correlated Gaussian variables with correlation
matrix $\hat C$ is obtained from the set of $\{w_1,....,w_N\}$ independent
random Gaussian variables with zero mean and unitary variance by
applying the linear transformation $u = A w$, where $A$ is the lower
diagonal matrix $A$ such that $C = AA^t$ (Cholesky decomposition of
the matrix C).  
\item The second method makes use of discrete Fast Fourier
Transform (FFT) to generate the stationary Gaussian field
$\{u(x_n)\}$ over a grid of sites $x_n$ with $n=1,...,{\cal N}$.  
The values of the field on the site $x_n$ is expressed as the Fourier sum
$$
u(x_n) = \frac{1}{\cal N} \sum_{k} \mbox{e}^{ik x_n} z(k)
$$ with $k=2\pi m/{\cal N}$ and $m=0,1,...,{\cal N}-1$ to avoid
aliasing.  Note that $z(k) = z^*(-k)$ to ensure that $u(x_n)$ is real.
The choice $z(k) =[\alpha(k) + i\beta(k)]/2$ with $\alpha(k)$ and 
$\beta(k)$ independent real Gaussian variables of zero mean and variance 
${\cal N} \hat c(k)/2$, guarantees the set of random variables $\{u(x_n)\}$
representing the discretized version of the field to have the correct
power spectrum $\hat c(k)$.
\end{enumerate}
Once the fields is generated the position of a 
particle $i$ is updated according to the Euler scheme 
\be
x_i(t+h) = x_i(t) +  \sqrt{h} u(x_i) 
\label{eq:move_h}
\ee with time step $h$.  When the Gaussian displacement field is
generated through FFT, there is clearly a problem 
associated to discretization due to the grid of step $\Delta x$
where the FFT is computed. We assign a particle the displacement $u_n$
if its position at time $t$ falls in the $n$-th bin $[n\Delta
x,(n+1)\Delta x]$ determined by the grid.  The choice of the time step
is such that $h \langle u(x_i)^2 \rangle \simeq \Delta x^2$ in order
to sample the maximal resolution scale allowed by the discretization.

\begin{figure}
\includegraphics[clip=true,keepaspectratio,width=8.cm]{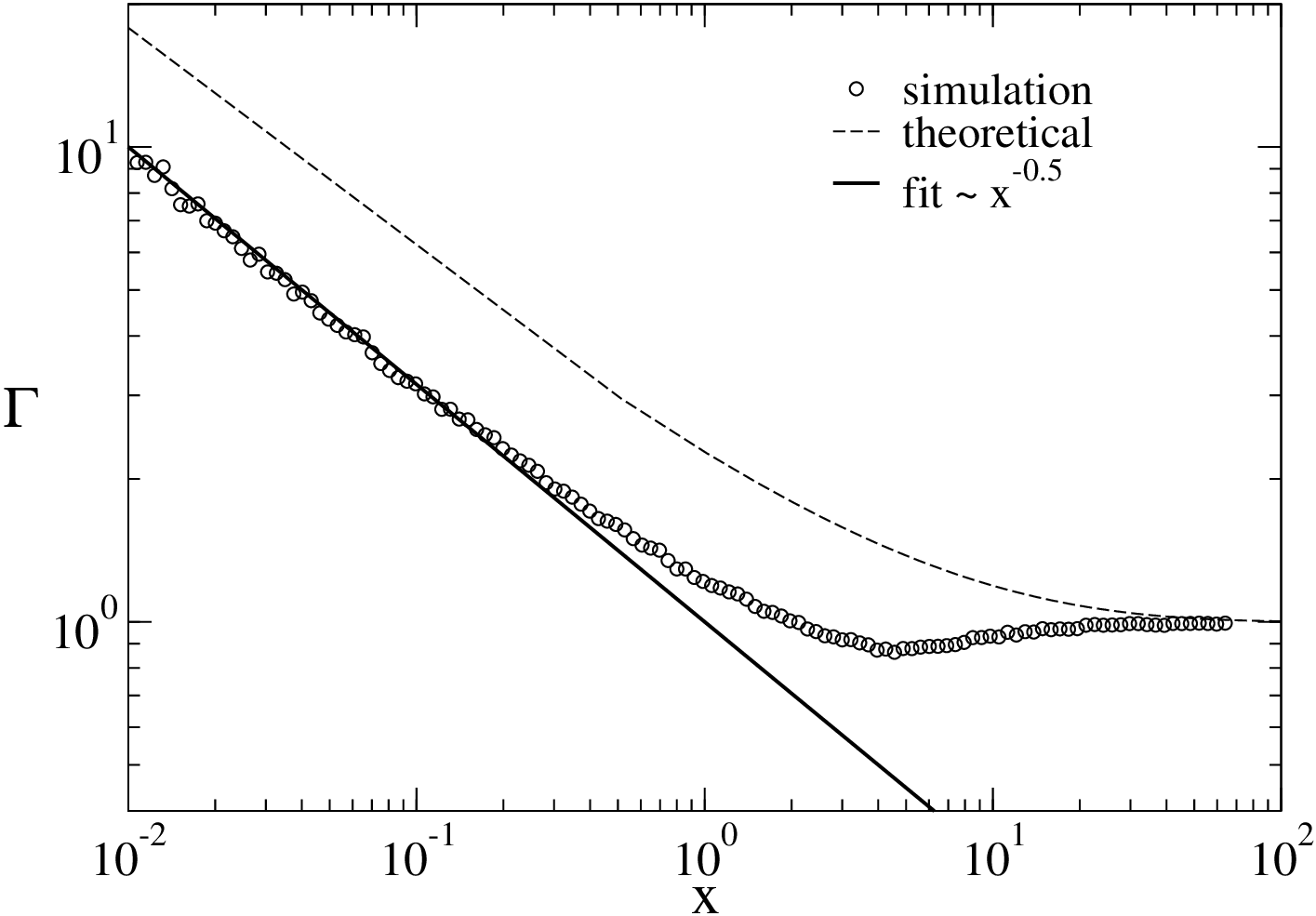}\hspace{0.cm}
\includegraphics[clip=true,keepaspectratio,width=8.cm]{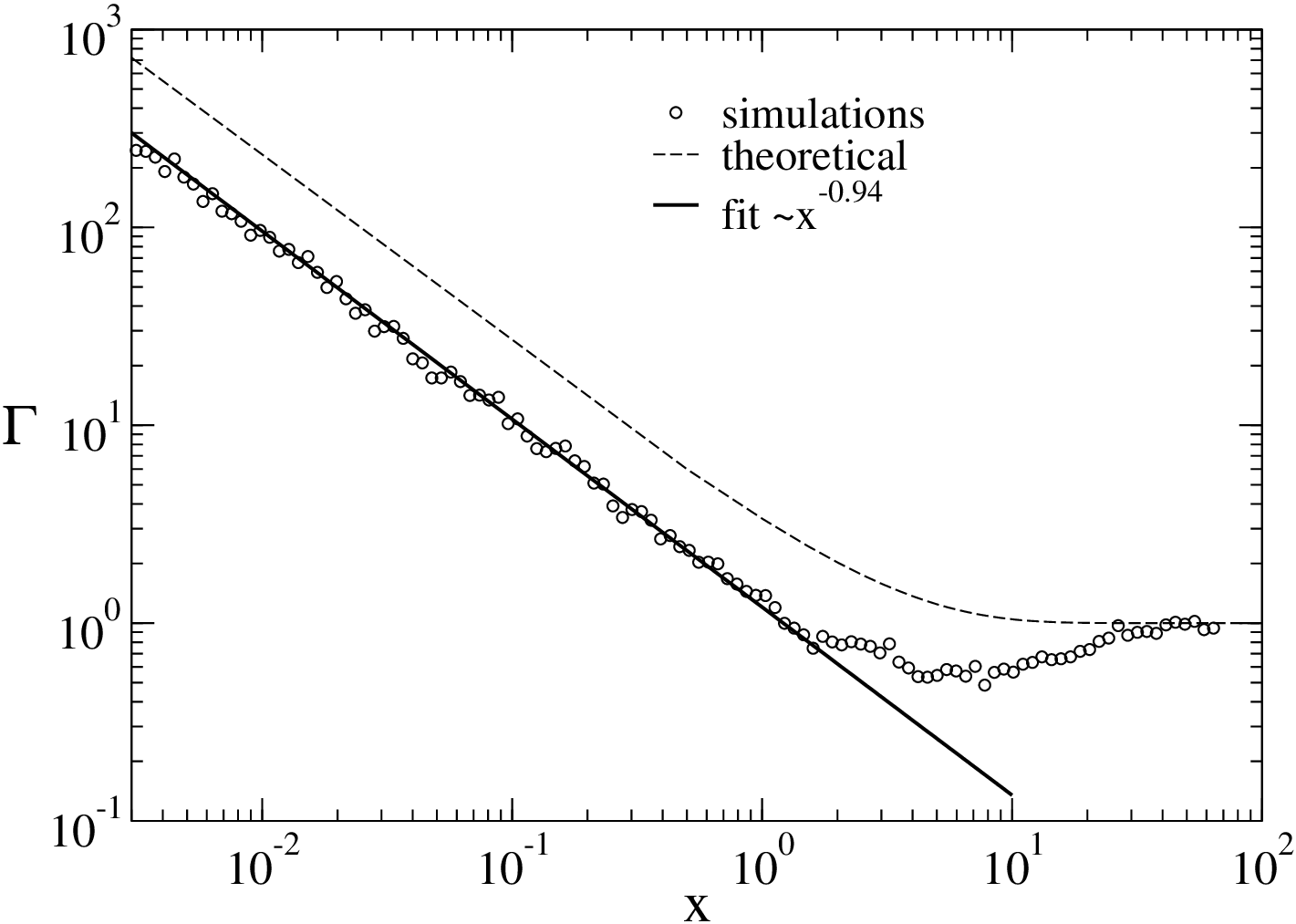}
\caption{Log-log behavior of the average conditional particle density
$\Gamma(x)$ as function of the separation $x$ generated by the
iterations of Eq.~(\ref{eq:move_h}) with time step $h=10^{-4}$ and a
displacement field with correlation (\ref{eq:corrfield}) defined by
parameters $x_0=3$, $\xi=0.5$ (Left) and $\xi=0.95$ (Right).
$\Gamma(x)$ is computed as the histogram of the inter-particle
distance binned exponentially in $128$ intervals.  The number of
particles is $N=256$ and simulation data are the results of the
average over $1500$ independent runs.  The small scale decay is
expected to be a power law with exponent $\xi$.  The solid line
indicates the result of a power law fitting with exponents $-0.5$,
$-0.95$ respectively and the dashed lines refer to Eq.~(\ref{ex2}).}
\label{fig:gamma0.5}
\end{figure}

\begin{figure}
\centerline{
\includegraphics[clip=true,keepaspectratio,width=8.cm]{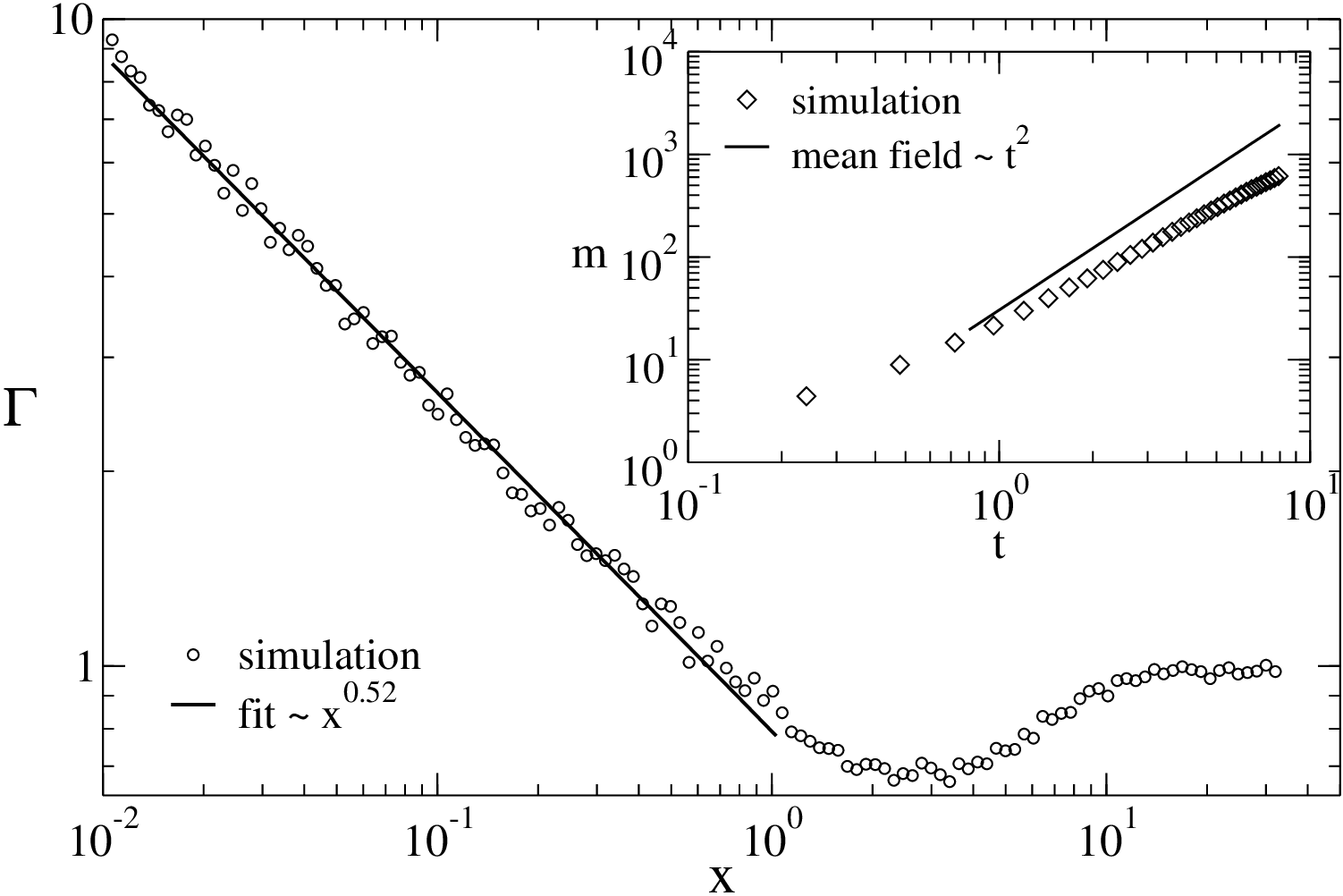}
}
\caption{Log-log plot of the average conditional density $\Gamma(x,t)$ 
as function of the separation 
$x$ at time $t=8$ , for $\xi=1.5$, $x_0=3$ and $N=64$ particles, 
averaged over $2200$ independent runs. Integration time step in 
Eq.~(\ref{eq:move_h}) amounts to $h=10^{-5}$.    
The small scale decay is fitted by a power law (solid line) 
with exponent $0.52$ in 
consistent with the theoretical value $(1-\xi)=0.5$. 
The inset shows the time increasing behavior of the coefficient of the 
$\delta$ function contribution to Eq.~(\ref{eq:gamma-x-t}) due to 
particle coalescence. The solid line indicates the mean field scaling 
behavior $\sim t^{1/(2-\xi)}$.}
\label{fig:gamma1.5}
\end{figure}

Spatio-temporal patterns obtained via simulations of a system of $N$ 
particles subjected to displacements of correlation $c(x)$ with $x_0=3$
are shown in Fig.~\ref{fig:evol} for cases $\xi=0.5$ and $\xi=1.5$.
The particle are initially distributed uniformely 
in a 1-dimensional simulation box with density $\rho=1$, periodic boundary 
conditions are applied at the ends of the box. 
For $\xi=0.5$ (left panel) and assuming elastic collisions (a reflecting 
boundary condition when particle trajectories intersect one another), 
the system exhibits simple particle clustering toward a stationary state.  
Instead for $\xi=1.5$ (right panel), particle
trajectories coalesce more and more in time and no stationary regime
is actually reached. 
We measured during each run the density correlation function $\Gamma(x)$, 
as the
histogram of the relative particle distance $|x_i-x_j|$ and the
results, averaged over several independent runs starting from the
uniform particle distributions, are plotted in
Figs.~\ref{fig:gamma0.5} and \ref{fig:gamma1.5}, for
$\xi=0.5, 0.95, 1.5$ respectively.
The stationary state for the case $\xi=0.5$
coincides with that described by theoretical Eq.~(\ref{eq:1d-st})
which is represented in Fig.~\ref{fig:gamma0.5} (together with the one
of the case $\xi=0.95$ which is in the same class of behavior). The
discrepancy in the amplitudes between numerical and theoretical
stationary $\Gamma_s(x)$ is due to the following finite size effect:
in simulations of of $N$ particles in a volume $V$, the average
conditional density $\Gamma(x)$ is subjected to the integral
constraint $\int_V \Gamma(x) dx =(N-1)$. Consequently, if the dynamics
develops a strong positive density correlation at small scale, an
artificial negative correlation must appear at larger scales.  The
typical behavior of $\Gamma(x,t)$ for $\xi=1.5$ is represented in
Fig.~\ref{fig:gamma1.5}. It shows correctly the small $x$ behavior
proportional to $x^{1-\xi}$ predicted by theoretical results.


\section{Conclusions and Discussion} \label{conclusion}
The evolution of several many particle system can be described
and also generated via the iterated application on the particles of
suitable displacement field which determines their trajectories.  In this
context, the question we addressed concerns the connection between the
statistical properties of the displacement field at a single time and
the spatial correlations that arises in the particle distribution
during the evolution.  We have shown how to derive the partial
differential equations describing the continuous time evolution of the
two-point correlation function of the particle density under the
iterative application of a stochastic displacement field with no temporal
memory from the corresponding exact single step evolution equation.
This continuous time equations are of a diffusion type and 
describe simple, fractional, or spatially correlated diffusion of
density fluctuations, depending only on the two-point statistical
properties of the elementary displacement field.  
Simple and
fractional diffusion occurs in absence of spatial correlations of 
displacements. Which of the two regime arises depends only on the 
finiteness of the elementary displacement variance.  
Spatially correlated displacements, instead, 
determine a Fokker-Planck (FP) equation for density correlations 
equivalent to a simple Langevin equation whose multiplicative noise
is defined again by the two-point displacement
correlation function.  Such a FP equation predicts a rich
phenomenology, ranging from simple particle clustering to
coalescence. These two regimes depends uniquely on the small scale
behavior of displacement correlations.

We characterized and classified these regimes in terms of the
the solutions $\Gamma(x,t)$ of the FP equation for the
density correlations which depend on the small scale statistical
properties of the displacements fields.  This classification can be
given in terms of the theory of boundary conditions of FP equation
\cite{Vankampen}. Our problem, indeed, presents in general a
singularity at the origin corresponding to the vanishing of the
structure function of the velocity field and which has to be treated
as a possible additional boundary \cite{Gardiner}.  
By developing a simple
theoretical approach in $d=1$, we have compared the results with the
solution classification for the Kraichnan ensemble for 
$d-$dimensional turbulence \cite{verg-gaw} finding a perfect
agreement.

Computer simulations, implementing the evolution of a one dimensional 
system of particles driven by a Gaussian correlated displacement field, 
confirm the presence of the clustering and coalescence regimes in agreement 
with our theoretical predictions.

It is noteworthy that the problem we have considered is different from
the motion of particles in random potentials for which in general
different FP equations hold \cite{Vankampen}. This point is simply
clarified by the observation that, in our model, for any spatial
correlation of the displacement field, any single particle performs a
simple Brownian motion, and the effect of the two-point displacement
correlations amounts to correlating the Brownian trajectories of different
particles.  On the contrary in presence of an external random
potential even the motion of a single particle is far from being
simply Brownian.

It is also interesting to discuss the analogies with the random 
one dimensional coagulation process $A+A \to A$ \cite{Coagul}, 
where particles perform mutually independent 
random walks and when they collide, one of them is removed according to 
an assigned rule. The particles are considered
to be ``sticky'' as they undergo perfectly inelastic collision 
and coalesce in a single particle independently of the relative velocities.  
In our case on the contrary, for $1<\xi<2$,
particles are not ``sticky'' and their coalescence is only a consequence
of the small scale correlation properties of the displacement field
which imply particle-particle collisions occurring at vanishing
relative velocity. Therefore, even though we would impose
perfectly elastic collisions, particles would coalesce the same.
The process $A+A\to A$ is recovered in the limit $\xi\to 0$ 
(implying collisions at non-zero relative velocity with probability one) 
and imposing inelastic collisions at interparticle contact.

In summary, via simple physical arguments, we have generalized the result of
Ref.~\cite{coha}, concerning only the smooth case $\xi = 2$, to the case
of $0 <\xi <2$ where $\xi$ is the exponent defining the behavior of
displacement correlation $c(x)$ at short scales: 
$d(x)=[c(0)-c(x)] \sim x^{\xi}$.

 

\end{document}